\documentclass[useAMS,usenatbib,referee]{raa}
\usepackage[colorlinks=true,citecolor=blue]{hyperref}
\usepackage{graphicx,times}             
\input{psfig.sty} 
\usepackage{epsfig}
\usepackage{natbib}
\newcommand{\Msun}{M$_{\odot}$}

\begin{document}

\title{A multiwavelength view of the ISM in the merger remnant Fornax A galaxy}

\volnopage{Vol.0 (200x) No.0, 000--000}      
\setcounter{page}{1}          

\author{S. P. Deshmukh \inst{1}, B. T. Tate \inst{2}, N. D. Vagshette \inst{3}, S. K. Pandey \inst{4}, \and M. K. Patil \inst{3,*} }

\institute{Department of Physics, Institute of Science, Nagpur 440 008, India \\ \and
      Department of Physics, Balbhim Arts, Science and Commerce College, Beed 431 122, India \\ \and
  School of Physical Sciences, S. R. T. M. University, Nanded 431 606, India \\
          \hspace{0.4cm}E-mail: {\it patil@iucaa.ernet.in}\\
           \and
           School of Studies in Physics \& Astrophysics, Pt. R.S. University, Raipur 491 010, India \\
           }

   \date{Received~~0000 month day; accepted~~0000~~month day}

\abstract{We present multi-wavelength imagery of the merger remnant galaxy NGC 1316 with an objective to study the content of dust and its association with the other phases of the ISM. Color-index maps as well as extinction maps derived for this galaxy reveal an intricate and complex dust morphology in NGC 1316, i.e., in the inner part it exists in the form of a prominent lane while at about 6--7\,kpc it apparently takes an arc-like pattern extended along the North-East direction. In addition to this, several other dust clumps and knots are also evident in this galaxy. Dust emission mapped using \textit{Spitzer} data at 8 $\mu$m indicates even more complex morphological structures of the dust in NGC1316. The extinction curve derived over the optical to near-IR bands closely follows the standard Galactic curve suggesting similar properties of the dust grains. The dust content of NGC 1316 estimated from optical extinction is $\sim$ 2.13$\times\, 10^5$ \Msun. This is a lower limit compared to that estimated using the IRAS flux densities $\sim$ 5.17$\times\, 10^6$ \Msun\, and flux densities at 24$\mu$m, 70$\mu$m and 160 $\mu$m from MIPS $\sim$ 3.2$\times\, 10^7$ \Msun. High resolution \textit{Chandra} observations of this merger remnant system have provided with an unprecedented view of the complex nature of hot gas distribution in NGC 1316 which closely matches the morphology of ionized gas and to some extent with the dust also. X-ray color-color plot for the resolved sources within optical D$_{25}$ extent of NGC 1316 has enabled us to separate them in different classes.
}
\authorrunning{Deshmukh et al.} 
\titlerunning{Multiphase ISM in the Fornex A (NGC 1316)} 

   \maketitle


\section{Introduction}
Multi-wavelength data acquired on the early-type galaxies (ETGs) using ground based as well as space telescopes have greatly enhanced our understanding regarding the  origin of multiphase ISM in this class of galaxies \citep{1994A&AS..105..341G, 2005A&A...433..497R, 2007A&A...461..103P, 2011MNRAS.416.1680C, 2010MNRAS.409..727F,2012MNRAS.422.1384F}. The general picture that emerges from the past studies is that either the cold/warm ISM observed in these galaxies is of external origin, accreted through merging and/or close-encounter episodes with neighboring galaxies, or is the by-product of cooling of the hot ISM that has originated from stellar mass loss \citep{2003ApJ...586..826K}. Systematic study of these galaxies selected from different environments delineate that relative contribution of the two competitive processes i.e., external versus internal, do not follow a general rule regarding their origin but varies greatly among the objects. Several attempts have been made to probe properties and origin of the ISM in early-type galaxies using spectroscopic observations studying dynamics and chemical abundance of the gas and stars \citep{1987IAUS..127..135B,1989ARA&A..27..235K,2006MNRAS.366.1151S,2010A&A...519A..40A},however, due to the limited spectral coverage of the study and limited knowledge of the interplay between different phases of ISM, we could not arrive at a decisive conclusion regarding the true nature of the ISM. Therefore, multi-wavelength study of various components of ISM in a large sample of E/S0 galaxies with dust lanes is important to investigate the nature and origin of ISM in this class of galaxies.  

Unusual optical signatures discernible in the form of dust-lanes, shells, tidal features, double nuclei, etc. are the direct evidences indicative of the merger like events that the host galaxies might have experienced in the past \citep{1987IAUS..127..135B, 2004ApJ...613L.121G,2007A&A...461..103P}. Another evidence regarding external origin of the dust is provided by the amount of dust content of such galaxies. It is found that the observed amount of dust using the optical extinction measurement and the IRAS flux densities is always larger by several factors than that expected from the mass loss of evolved stars \citep{2004ApJ...613L.121G,1999AJ....118..785D,2007A&A...461..103P}. Furthermore, kinematical studies of dust lane early-type galaxies have revealed that the motion and orientation of the gas in many systems is decoupled from the stellar rotation, and hence provides an additional evidence for their external origin \citep{1987IAUS..127..135B,1989ARA&A..27..235K,2001ApSSS.277..409C}. If the dust is originated internally through the evolution of single stellar population, then instead of forming disks, lanes, etc., it would be evenly distributed throughout the galaxy \citep{2012MNRAS.422.1384F}.

NGC 1316 is a peculiar S0 galaxy, with numerous tidal tails, shells and many pronounced dust patches including the prominent dust lane oriented along its optical minor axis. In addition to dust, NGC 1316 also host H$\alpha$ filaments, strong shells \citep{1983ApJ...272L...5M} and several filaments and loops of ISM \citep{1988ApJ...328...88S}. All these features confirm that NGC 1316 have experienced a strong merger like episode in the past \citep{2002MNRAS.330..547T}. X-ray observation of this galaxy with \textit{Einstein} \citep{1992ApJS...80..531F}, \textit{ROSAT} PSPC \citep{1995ApJ...449L.149F} and \textit{CHANDRA} \citep{2003ApJ...586..826K} and their systemic analysis have confirmed a low luminosity AGN with 0.3-8.0 keV L$_X \sim 5.0 \times 10 ^{39}$ erg s$^{-1}$ \citep{2003ApJ...586..826K} and also exhibit extended X ray emission with inherent several substructures. NGC 1316 is one among the nearest giant radio galaxy with well defined core-jet-lobe structure. The lobes of this galaxy shows smooth light distribution and sharp boundaries. The bridge of emission in this galaxy is found to be significantly displaced from its centre, providing additional evidence for the strong merger episode \citep{1983A&A...127..361E}. \cite{2001MNRAS.328..237G} based on the study of bright globular clusters within NGC 1316 have estimated age of this merger remnant to be $\sim$3\,Gyr. Thus, NGC 1316 suits a potential candidate to investigate properties of dust and other phases of ISM.  

In this paper we present multicolour CCD imaging involving broad band B V R I and narrow band H$_{\alpha}$ data. We mainly focus on investigating the distribution of dust its extinction properties and compare its association with the ionized gas. We also present reanalysis of \textit{Chandra} observations of NGC 1316. The paper is organized as follows. In section 2 we describe the optical and X-ray observations and the data analysis process. Section 3 discusses the dust extinction properties and morphology of ionized gas. Results derived from the spatial and spectral analysis of X-ray photons along with the discrete sources are also discussed in this section. Section 4 discusses the issue of origin of dust as well as its association with other phases of ISM. Finally, we summarize our results in section 5. We adopt optical luminosity distance of 25 Mpc throughout this analysis.
\begin{figure}
\centering
\includegraphics[width=100mm,height=100mm]{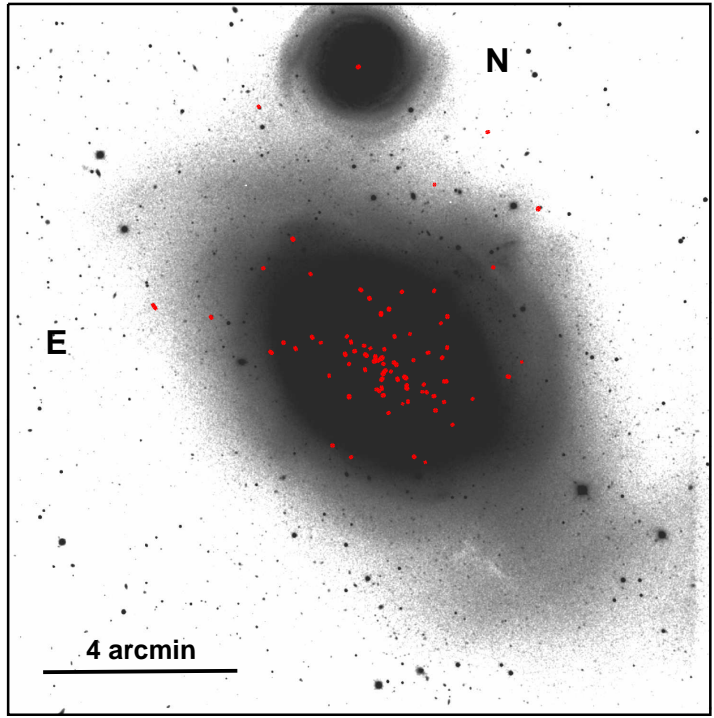}
\caption{\label{raw} Cleaned, background subtracted B-band image of NGC 1316, overlaid on which are the \textit{Chandra} X-ray point sources detected within ACIS-S3 chip. This figure clearly reveals a set of shells, ripples as well as tails; signatures of merger remnant}
\end{figure}

\begin{table}
\begin{center}
\label{symbols}
\caption{Global parameters of NGC 1316.}
\begin{tabular}{llllllllll}
\hline
   Parameter & Value \\     
\hline
 Alternate names & Fornax A; IRAS 03208-3723; PGC 012651 \\
    		 & ESO 357-G 022; NVSS J032241-371225     \\
 RA; DEC         & 03:22:41.7; -37:12:30           \\
 Morphology      & SAB0(s)pec / S0                       \\
 Mag B$_T$       & 9.42                              \\
 D$_{25}$         & 12'.0 x 8'.5                       \\
 Redshift (z)     & 0.00587                          \\
 Effective radius (kpc) & 7.08    \\
 IR flux densities (Jy) & 0.33$ \pm $0.04 (12$\, \mu$m); 3.07$\pm$0.03 (60$\, \mu$m); 8.11$\pm$1.99 (100$\, \mu$m) IRAS \& \\
                     & 0.43$ \pm $0.02 (24$\, \mu$m); 5.44$\pm$0.40(70$\, \mu$m); 12.61$\pm$1.78 (160$\, \mu$m)  MIPS  \\
\hline
\end{tabular}
\end{center}
\end{table}

\begin{figure*}
\includegraphics[width=17.5cm,height=8.5cm]{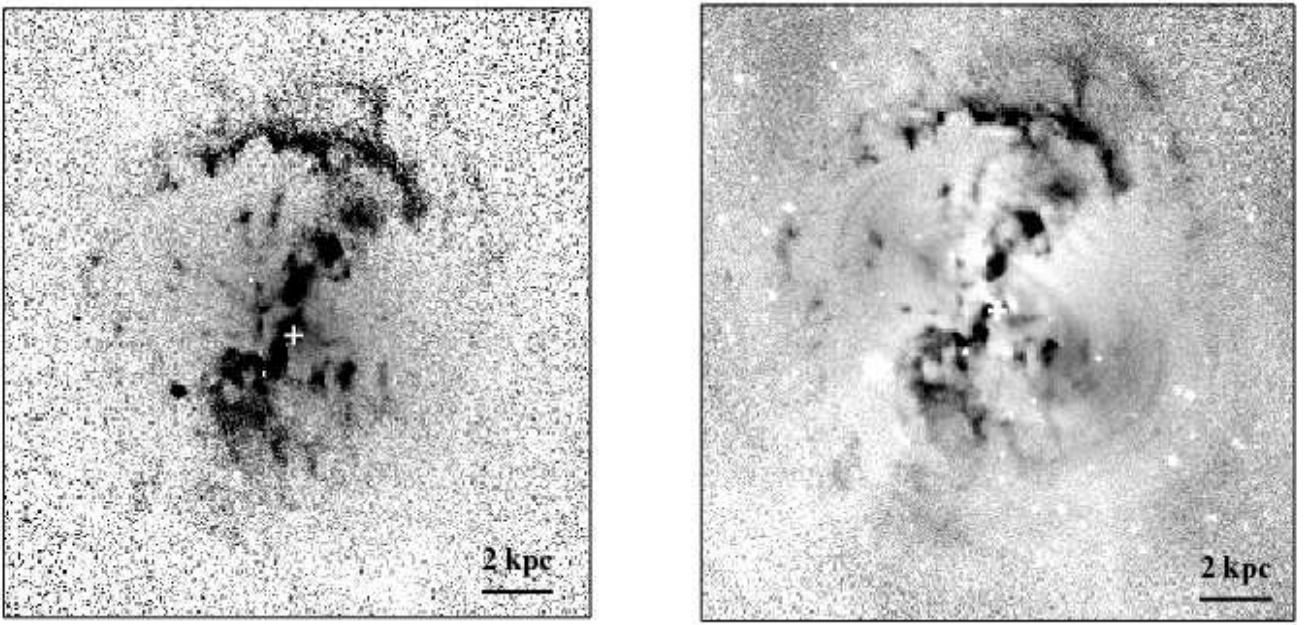}
\caption{\label{dustm} Central $2'.5\times 2'.5$ region of NGC 1316, \emph{(left panel)} (B-V) colour index map, darker shades delineates the dust occupied regions. Cross indicates position of the galaxy center as recognized by NED. ({\it right panel}) B-band extinction map of NGC 1316.}
\end{figure*}

\section[]{Observations and data preparation}
\subsection{Optical and near-IR data}
Deep, high S/N CCD images on NGC 1316 in \textit{B, V, R, I, and H$_{\alpha}$} filters were acquired from NED that were observed  with the Cerro Tololo Inter-American Observatory (\textit{CTIO}) / University of Michigan Curtis Schmidt 0.6/0.9 m telescope (see \citealt{1998AJ....115..514M} for details). The detector was a Thomson 1024 $\times$ 1024 CCD, pixel size of 19$\mu$m and spatial resolution of 1\arcsec.83 pixel$^{-1}$. These images were pre-processed i.e., bias-subtracted, flat-fielded, co-added and exposure corrected \citep{1998AJ....115..514M}. The sky background was estimated using the box method \citep{1998A&A...333..803S} and was then subtracted from the respective passband image. The geometrically aligned, background subtracted images were then used for the dust extinction study in the target galaxy. One of such cleaned B band image is shown in Figure~\ref{raw}, which reveals several shells, ridges and ripples around the main galaxy at larger radii, indicative of the merger signatures. With an objective to extend the study of dust extinction over other passbands, we acquired near-IR J, H \& Ks band images on NGC 1316 from the archive of Two Micron All Sky Survey (2MASS) observatory. 
\subsection{X ray data}
Though X-ray emission properties of NGC 1316 have already been reported using the observations from \emph{Chandra} \citep{2003ApJ...586..826K,2010ApJ...721.1702L}, \emph{XMM-Newton} \citep{2006ApJ...645..256I,2010ApJ...721.1702L}, \emph{SUZAKU} \citep{2010PASJ...62.1435K}, \emph{Einstein} \citep{1992ApJS...80..531F},  as well as \emph{ROSAT} \citep{1998ApJ...497..699K}, however, with an objective to examine association of hot gas with dust and ionized gas in this merger remnant galaxy, we have made use of high resolution X-ray data on NGC 1316 from the archive of \emph{Chandra} observatory. NGC 1316 was observed by \emph{Chandra} on 2001-04-17 (Obs. ID 2022) with the ACIS-S3 as the aimpoint for an effective exposure of 30.0 ks. 

Standard tasks available within \textit{Chandra} Interactive Analysis of Observations (CIAO version 4.2.0) and recent calibration files provided by the \textit{Chandra} X-ray Centre (CXC) (CALDB version 4.3.0) were employed for the analysis of X-ray data. These data sets were first filtered for the periods of high background emission using 3$\sigma$ clipping of the 0.3 - 10.0\,keV light curve extracted from the chip with a binning of 260\,s. This resulted in to the net exposure time of 25.0\,ks. For background subtraction we used properly scaled blank sky background files provided by the CXC. Point sources recorded on chip S3 were detected using the \emph{wavdetect} tool within CIAO adopting a detection threshold of 10$^{-6}$ and scale parameter covering 6 steps between 1 to 32 pixels. This has enabled us to detect a total of 86 discrete sources within the S3 chip. Out of the 86 detected sources, 80 were lying within the optical D$_{25}$ region of NGC 1316 (Figure~\ref{raw}).  
\section{Results}
\subsection{Dust properties}
\subsubsection{Dust Extinction}
Though NGC 1316 is known to host dust features since their first time detection by \citep{1980ApJ...237..303S}, quantitative analysis of extinction properties of dust in this galaxy are not available in the literature. To investigate the amount of dust extinction and its wavelength dependent nature, it is required to know the spatial distribution and extent of the dust in target galaxy. For this we generated color index maps (B-V), (B-R), (B-I), etc. of NGC 1316 by comparing light distribution in the geometrically aligned, seeing matched broad band images in different passbands. Figure~\ref{dustm} (\textit{left panel}) shows one of such (B-V) color-index map of NGC 1316, where darker shades represent the dust occupied regions and are consistent with those reported by \cite{1980ApJ...237..303S}. This figure reveals a prominent dust lane along optical minor axis of NGC 1316, which then takes an arc like shape at about 6-7 kpc. In addition to these main features, several filament and clump like features are also evident in this figure.

Comparison of light distribution in extinguished part of the galaxies with that in the absence of dust allows us to investigate extinction properties of dust and its wavelength dependent nature \citep{1987MNRAS.225..257B,2007A&A...461..103P}. This can be done by deriving dust free smooth models of the target galaxy in different passbands. Here, the dust free models of NGC 1316 were generated by fitting ellipses to the isophotes in optical broadband images using the ISOPHOTE package within IRAF (see \citealt{2007A&A...461..103P} for details). The position angle, ellipticity and center coordinates were kept free during this fit till the signal reaches 3$\sigma$ of the background. Regions occupied by dust and foreground stars, as evident in the colour index maps, were masked and ignored during the fit. Dust free models thus generated were used to quantify the wavelength dependent nature of dust extinction, \emph{extinction curve}, using the relation \citep{2007A&A...461..103P}
\hspace{10mm} \[A_{\lambda}=-2.5\,{\log \left(\frac{I_{\lambda,obs}} {I_{\lambda,model}}\right)}\]
where A$ _{\lambda} $ is the amount of total extinction in a particular passband (\textit{B,V,R,I}), while I$ _{\lambda,obs} $ and I$ _{\lambda,model} $ are the observed (attenuated) and un-extinguished (modeled) light intensities in a given passband, respectively. One of such extinction maps derived for NGC 1316 is shown in Figure~\ref{dustm} (\emph{right panel}). This figure confirms the unusual, intriguing and clumpy morphology of dust, as was evident in (B-V) colour index map of NGC 1316. From this figure it is apparent that in the inner part, dust appears in a well-defined lane oriented along its optical minor axis which then takes an arc like form at about 6-7 kpc  oriented along the North-East direction. In addition to these main features, several knots, dust patches and clumps are also evident in this figure.

\begin{figure}
\centering
\includegraphics[width=90mm]{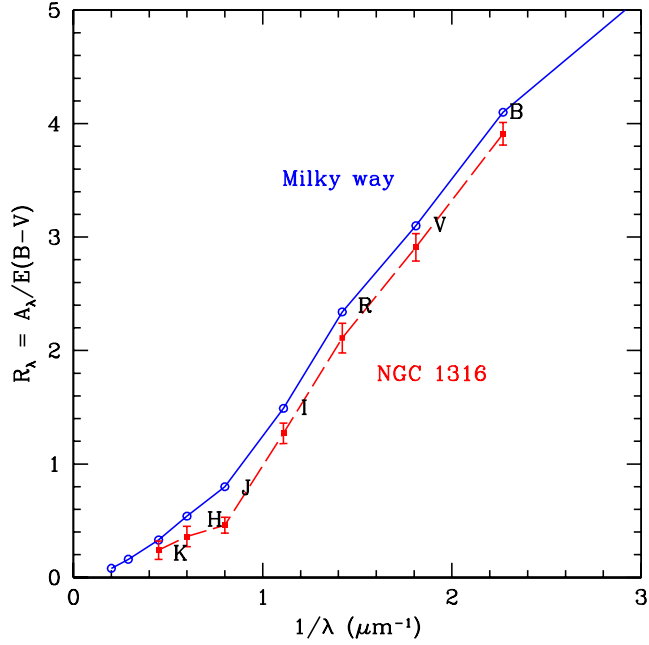}
\caption{\label{ext} Optical through near-IR extinction curve (dotted line) for NGC 1316 plotted as a function of inverse of the wavelength. For comparison we also plot the standard Galactic extinction curve (solid line)}
\end{figure}

To investigate quantitative properties of dust extinction and to examine its wavelength dependent nature, total extinction values were measured in each pass-band by sliding a 5$\times$5 box on the dust occupied region. Numerical values of local extinctions in each pass bands (A$ _{\lambda}$) were then used to derive the extinction values R$_{\lambda}$ by fitting linear regressions between the total extinction A$_\lambda$ and the selective extinction $E(B-V)=A_B - A_V$. The best-fitting slopes of these regressions along with their associated uncertainties were subsequently used to derive the R$_\lambda$ and hence the extinction curve. The  extinction curve derived over optical-to-near-IR region of the electromagnetic spectrum for the dust occupied regions in Fornax A (NGC 1316) galaxy is shown in Figure~\ref{ext}. From this figure it is clear that the dust extinction curve plotted between the measured values of R$_\lambda \left(=\frac{A_\lambda}{E(B-V)}\right)$ versus inverse of wavelength of their measurement varies linearly and follows closely the standard Galactic extinction law \citep{1977ApJ...217..425M}; consistent with those derived for several other external galaxies \citep{1994A&AS..105..341G,1996A&A...314..721S,2001BASI...29..453P,2002BASI...30..759P,2003asdu.confE..16P,2007A&A...461..103P,2009arXiv0901.1747P,1999AJ....118..785D,2012MNRAS.422.1384F,2012NewA...17..524V}. Relatively smaller value of R$_{\lambda}$ in the case of NGC 1316 imply that, the dust grains responsible for the extinction of optical and near-IR light are smaller than the canonical grains in the Milky Way $<a>\,\,<\, <a_{Gal}> $. This leads to R$_{V}$ = 2.91$\pm$0.12 relative to 3.1 for the Milky Way.
\subsubsection{Dust mass estimation}
Total extinction measured in V band was used to quantify the dust content of the NGC 1316. For this we integrated the dust column density $\Sigma_d$ over the image area (A) occupied by the dust features in V-band extinction map. Assuming that  chemical composition of the extragalactic dust is uniform throughout the galaxy and is identical to that in the Milky Way, we quantified the dust column density in NGC 1316. As the present study is limited to optical to near-IR bands, therefore, we employed the simple two-component model comprised of an adequate mixture of spherical silicate and graphite grains \citep{1977ApJ...217..425M}. This model assumes uncoated refractory particles with power-law size distribution. While the details of the dust estimation are described in \cite{2007A&A...461..103P}, the dust mass can be estimated by integrating the dust column density over the dust occupied area (A) using
\[ M_d = A \times \Sigma_d \]
For spherical grains of radius `$a$' distributed with number density `$n_d$' per unit volume in a cylindrical column of length `$l_d$' and unit cross-section area along the line of sight, the reduction in intensity at a given wavelength in units of magnitude i.e. total extinction at wavelength $\lambda$ is given by
\[A_{\lambda} = 1.086\,\, \pi a^2 Q_{ext}(a,\lambda) N_d\]
where $N_d$ = $\int n_d dl$ is the dust column density and $Q_{ext}(a,\lambda)$ is the extinction efficiency of the dust grain. 

Instead of spherical grains of constant radius `$a$', if we assume the size distribution of $n(a)$ $\propto$ $a^{-3.5}$ \citep{1977ApJ...217..425M} then the expression for total extinction at wavelength $\lambda$ becomes (please see \citealt{1994ApJ...422..164K}),
\[A_{\lambda} = 1.086\,\,l_d \int_{a_{min}} ^{a_{max}} \pi a^2\,\,Q_{ext}(a,\lambda)\,\,n(a)\,da = 1.086\,\,l_d \int_{a_{min}} ^{a_{max}} C_{ext}(a,\lambda)\,n(a)da\]
where $n(a)da$ represents number of grains per unit volume along the line of sight with radii in the range $a$ to $a+da$ (in cm$^{-3}$); $a_{min}$ and $a_{max}$ are the lower and upper cut-off of the grain size distribution (in cm); $l_d$ is the length of the dust column (in cm) and $C_{ext}(a,\lambda) =  \pi a^2\,\,Q_{ext}(a,\lambda)$ is the total extinction cross-section at wavelength $\lambda$ (in $cm^{2}$).

Using the value of $l_d$, the dust column density ($g\,\,cm^{-2}$) can be expressed as,
\[ \Sigma_d  = l_d \times \int_{a_{min}} ^{a_{max}} \frac{4}{3}\pi a^3 \rho_d\,\,n(a)\,da \]
This lead to the dust content of NGC 1316 to be equal to 2.13 $\times\, 10^{5}\, M _{\odot}$. For the average Galactic extinction curve with R$_V$=3.1, we assume $N_H/A_V\, = 1.87 \times 10^{21}\, cm^{-2}\, mag^{-1}$\, \citep{2003ARA&A..41..241D}. As this method assumes the screening effect of dust, therefore, is insensitive to the component of dust that is diffusely distributed throughout the galaxy. Thus, the dust mass estimated using optical extinction provides only the lower limit. Uncertainties involved due to the lower and upper cutoffs of the grain size may further worsen this estimate. 

The dust content of NGC 1316 can alternatively be estimated using the IRAS flux densities at 100$\mu$m and estimating the dust temperature by fitting single temperature  modified black body FIR SED over 60 - 500 $\mu m$ of emissivity proportional to $\lambda^{-1.5}$ \citep{2002ApJ...568...88Y,2011ApJ...738...89S}. In the present case the dust temperature is found to be T$_d$ = 26.8 K. Then we estimated the dust mass using the relation $M_d = \frac{D^2S_{\nu}}{\kappa_{\nu}B_{\nu}(T_d)}$, where $\kappa_{\nu}$ is the dust opacity, $S_\nu$ is the flux density, $B_{\nu}(T_d)$ is the Planck function for the dust grain temperature $T_d$ and $D$ is the distance of the galaxy in Mpc. Considering the value of dust emissivity given by \cite{1983QJRAS..24..267H}, at 100$\mu$m flux emission, the dust mass in solar units is given by \citep{1989ApJS...70..699Y},
\[ M_d = 4.78 S_{100\mu m}D^2\left[exp\left(\frac{143.88}{T_{dust}}\right)-1\right]\]
and is found to be equal to 5.17$\times$10$ ^{6}$ M$_{\odot}$; an order of magnitude higher than that estimated using optical extinction method. This discrepancy in the two estimates is due to the fact that optical extinction method is insensitive to the intermix component of the dust, while \textit{IRAS} can record this component. The estimate of dust temperature by a single temperature gray body SED fit over 60 - 500 $\mu m$ overestimates the dust grain temperature and hence underestimates the dust mass. Moreover, $\kappa \propto \lambda^{-1.5}$ assumes higher opacity at far-IR, therefore, leads to further underestimation of the dust content of the target galaxy. Therefore, the true dust content of the NGC 1316 was estimated using the integrated MIPS data at 24$\mu m$, 70$\mu m$ and 160$\mu m$ using the relation given by \cite{2009ApJ...701.1965M},
\[M_{dust} = \frac{4\pi D^2}{1.616 \times 10^{-13}} \times \left(\frac{<\nu S_{\nu}>_{70}}{<\nu S_{\nu}>_{100}}\right)^{-1.801} \times\, C\,\,\,\,\,\, M_{\odot} \]
where, $C=(1.559 <\nu S_{\nu}>_{24} + 0.7686 <\nu S_{\nu}>_{70} + 1.347 <\nu S_{\nu}>_{160})$, D is the distance in Mpc and $<\nu S_{\nu}>_{24}, <\nu S_{\nu}>_{70} and <\nu S_{\nu}>_{160}$ are the \textit{MIPS} flux densities at 24, 70 and 160 $\mu m$, respectively. The dust mass estimated using \textit{MIPS} flux densities is found to be equal to 3.21$\times$10$ ^{7}$ M$_{\odot}$ and is in agreement with that reported by \cite{2010ApJ...721.1702L} and \cite{2007ApJ...663..866D}, and is much higher than that estimated using the 100$\mu$m flux densities of IRAS.
\subsection{Association of Multiphase ISM}
\begin{figure*}
\includegraphics[width=16cm,height=8cm]{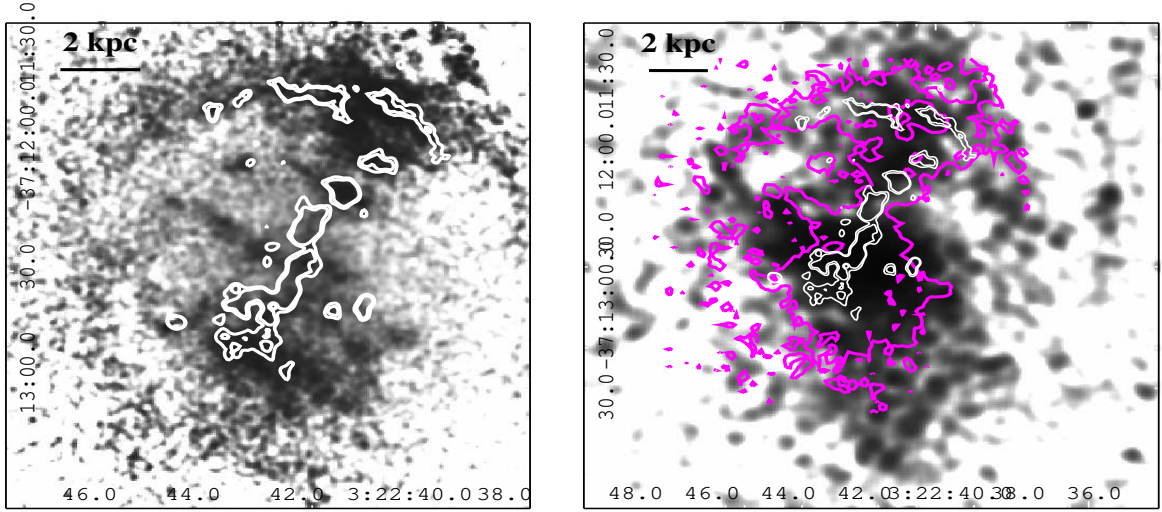} 
\vspace{-1.0cm}\caption{\label{ha} \emph{(left panel)}: Continuum subtracted H$\alpha$ emission map, overlaid on which are the dust extinction contours (white color). \emph{(right panel)}: 3$\sigma$ smoothed 0.3--3.0 keV X-ray emission map, overlaid on which are the H$\alpha$ emission contours (pink) and dust extinction contours (white).}
 \end{figure*}

With an objective to examine the association of dust with ionized gas, we have derived H$_\alpha$ + [N II] emission maps of NGC 1316 following the method discussed by \cite{1998AJ....115..514M}. This was done by subtracting the properly scaled, sky-subtracted R-band continuum image from that of the emission-line image. The adopted scale factor was calculated by carrying out the least-square fit to residuals of field stars in the two frames. The resultant spatial distribution of the ionized gas within NGC 1316 is shown in Figure~\ref{ha}(\textit{left panel}) and matches closely that of the dust in (B-V) colour map as  well as 3 $\sigma$ smoothed 0.3 - 3.0\, keV X-ray emission map (Figure~\ref{ha} right panel). The arc of ionized gas in the H$_{\alpha}$ emission map appears more prominently compared to that in the dust extinction map. 
\subsection{X-ray properties}
\subsubsection{Diffuse gas}
\begin{figure*}
\includegraphics[width=18cm,height=12cm]{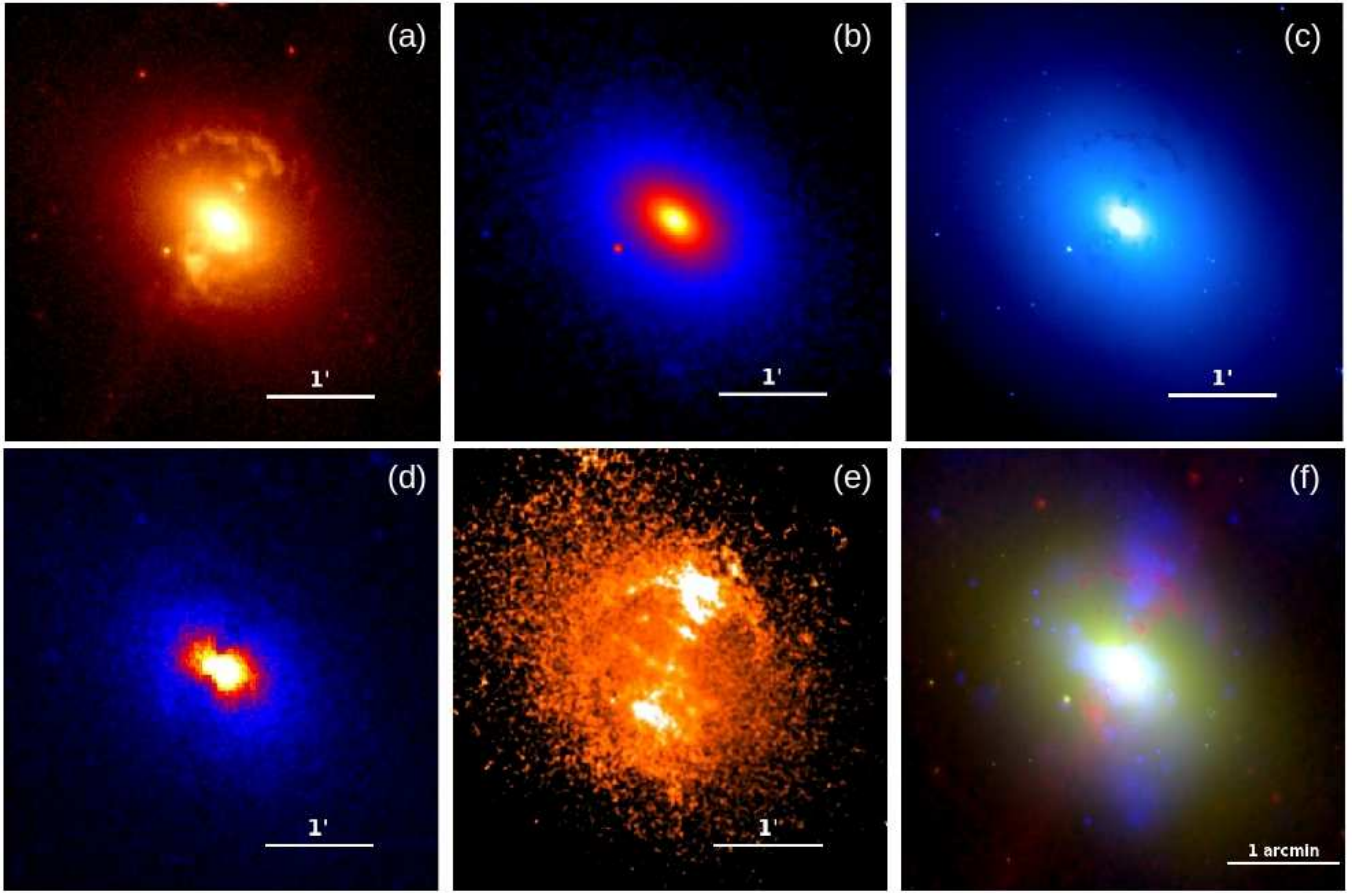}
\caption{\label{multi}Association of multiphase ISM in NGC 1316; \textit{(a)} Dust emission at 8$\mu$m detected by \textit{Spitzer}, \textit{(b)} stellar light distribution mapped through \textit{2MASS} K-band image,  \textit{(c)} stellar light distribution in B-band, \textit{(d)} Near-UV \textit{GALEX} image, \textit{(e)} continuum subtracted H$\alpha$ emission map and \textit{(f)} tri-color map delineating multiphase association; dust emission at 8$\mu$m represented in red, blue band light distribution in green and 0.3 - 3.0 keV hot gas in blue color.}
\end{figure*}

Diffuse X-ray emission map derived for a galaxy is the best tool to delineate morphology of hot gas. With an objective to examine association of dust and H$\alpha$ emitting ionized gas with that of the hot gas, we have derived 0.3 - 3.0\, keV X-ray emission map for NGC 1316 from the analysis of \emph{Chandra} observations and is shown in Figure~\ref{ha} (\emph{right panel}). This figure represents the background subtracted, exposure corrected, point source removed, 3$\sigma$ adaptively smoothed 0.3--3.0\,keV \emph{Chandra} image of NGC 1316 and is in agreement with that reported by \cite{1998ApJ...497..699K} using \textit{ROSAT} data and by \cite{2003ApJ...586..826K} using \emph{Chandra} data. Figure~\ref{ha} clearly reveals morphological similarities between the distribution of ionized gas and hot gas exhibiting a strong correspondence between the two. For comparison we overplot contours of the H$\alpha$ emitting gas on the X-ray image. Both these morphologies together closely follow that of the dust (Figure~\ref{ha} {\it right panel}) pointing towards common origin of all the three phases of ISM. X-ray emission in the energy band 0.3--3.0\,keV appears in more extended form compared to that of the ionized gas and shows a very disturbed structure along with filaments and patchy halos, perhaps X-ray cavities. In the inner region orientation of the dust lane appears to be slightly off those of the hot and warm gas. Figure~\ref{multi} illustrates the spatial correspondence between the different phases of ISM, including dust emission at 8$\mu$m from the \emph{Spitzer} data. Dust occupied regions mapped through the emission from PAH observed at 8$\mu$m using \emph{Spitzer} are found to coincide well with the X-ray emission in the central 7.5 kpc region.

To examine global properties of hot gas in the target galaxy, we have extracted a point source removed, background subtracted combined spectrum of the X-ray photons from within optical D$_{25}$ region of NGC 1316. To avoid contribution from the nuclear source we excluded central 20\arcsec region. The spectrum was fitted following the standard $\chi^2$ statistics within XSPEC version 12.6.0q with an absorbed single temperature MEKAL model. However, the fit exhibited residuals particularly in the higher energy range. Therefore, to constrain the emission from the unresolved sources we added a power law component to it. Even after adding power law component the fit was not reliable and exhibited residuals. Then we tried with a double temperature plus a power law components (mekal + mekal + pl), that resulted into a relatively better fit with $\chi^2$ value close to 1.62 for 83 \emph{dof}. During this fit, the hydrogen column density was fixed at the Galactic value of 2.40 $\times\, 10^{20}\, cm^{-2}$ \citep{1990ARA&A..28..215D}. The best-fit resulted in to kT$_{cool}$ = 0.18$\pm$0.03 keV, kT$_{hot}$ = 0.62$\pm$0.02 keV and the photon index equal to $\Gamma$=0.45$\pm$0.10. 

\begin{figure}
\centering
\includegraphics[width=8cm,height=7cm]{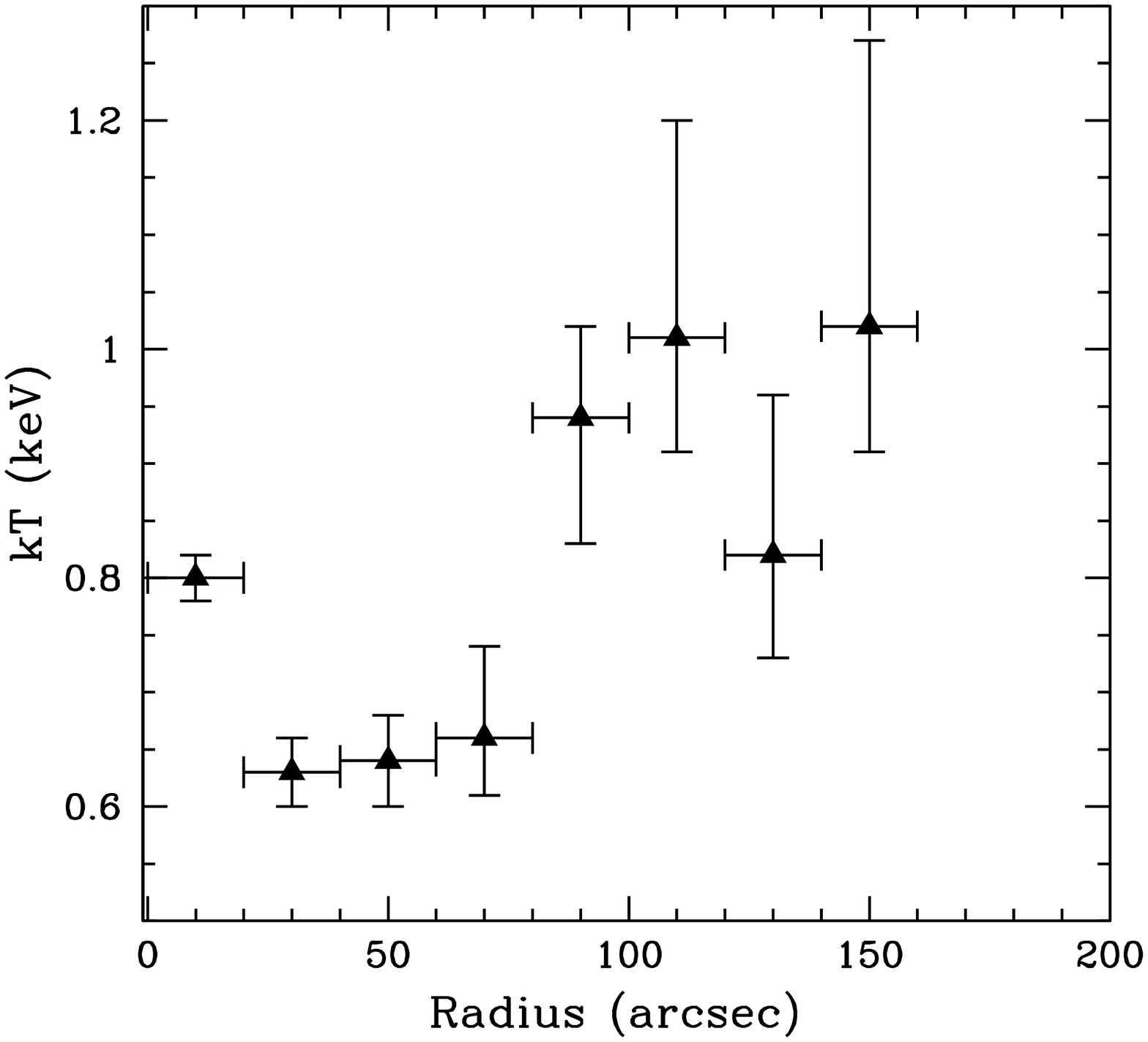}
\caption{\label{prof} Temperature profile of the hot gas distribution within NGC 1316 as a function radial distance. A positive temperature gradient can be confirmed from this figure.}
\end{figure}

With a view to examine temperature structure of the hot gas within NGC 1316, we performed spatially resolved spectral analysis of X-ray photons. For this we extracted 0.3--3.0\,keV X-ray photons from 8 different concentric annuli centered on the X-ray peak of the NGC 1316. Width of each of the annulus was set such as to get roughly same number of counts to validate the $\chi^2$ statistics. The background spectrum was extracted from the exposure corrected blank sky frame provided by the CXC. Source spectra, background spectra, photon-weighted response files and photon-weighted effective area files were generated for each of the annulus using the CIAO tool \textit{specextract}. Spectra extracted from each of the annulus were then fitted with a single temperature thermal plasma model (\textit{apec}) with neutral hydrogen column density fixed at the Galactic value \citep{1990ARA&A..28..215D}. Temperature, metal abundance and  normalization, etc. were kept free during the fit. The resultant radial temperature profile derived for NGC 1316 is shown in Figure~\ref{prof}. From this profile it is evident that the X-ray photons distributed within NGC 1316 shows temperature structure in the sense that temperature of the hot gas increases monotonically as a function of radial distance, like in the cooling flow galaxies \citep[and references there in]{2012MNRAS.421..808P}. A jump in the temperature profile is evident at about 1\arcmin.2 and is perhaps linked to the dust absorption.

\subsubsection{Discrete sources}
\begin{figure}
\centering
\includegraphics[width=9cm,height=8cm]{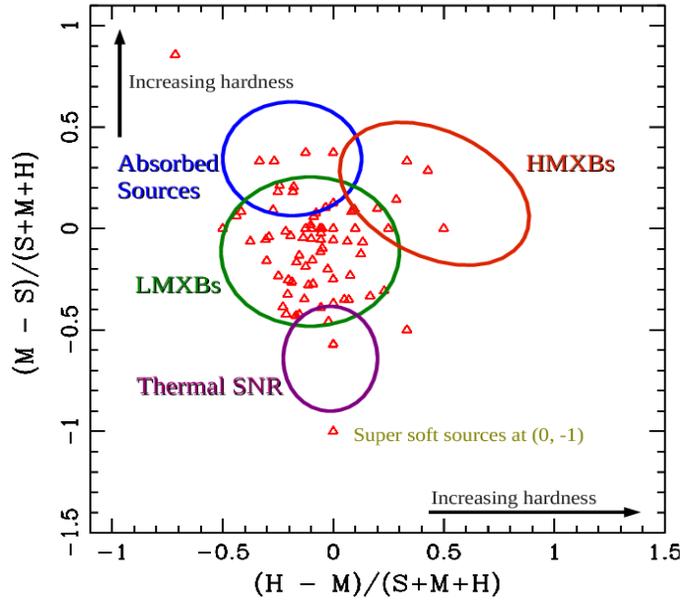}
\caption{\label{hr} X-ray color-color plot of the resolved sources plotted between X-ray hard color (H31) Vs. soft color (H21). Structural properties of the different classes of sources are highlighted in the figure.}
\end{figure}

As was discussed above, we detected 80 discrete X-ray sources within the optical D$_{25}$ region of this galaxy. Discrete sources within a galaxy are thought to be linked to the star formation history  and hence to the formation scenario of the host galaxy \citep{2003ApJ...587..356I, 2004ApJ...602..231C,2012arXiv1205.6057V}. X-ray color plot of the XRBs act as an efficient tool to investigate characteristic of individual sources \citep{2003ApJ...595..719P,2006ARA&A..44..323F}. Position of the source in this X-ray color plot clearly delineate its intrinsic nature and hence help us to classify them in different types \citep{2012arXiv1205.6057V}. To investigate X-ray characteristics of the resolved sources within NGC 1316, we derived X-ray color plot for these sources. For this, we extracted background subtracted X-ray photons from individual source in three different energy bands namely, soft (S, 0.3--1.0\,keV), medium (M, 1.0--2.0\,keV) and hard (H, 2.0--10.0\,keV), using the task \emph{dmextract} available within CIAO. Then, hardness ratio of individual source was estimated using the definitions $H21 = \frac{M-S}{S+M+H}$ and $H31= \frac{H-M}{S+M+H}$ \citep{2012arXiv1205.6057V}. The plot between X-ray hard (H31) versus soft (H21) color of all the sources is shown in Figure~\ref{hr}, from which it is apparent that majority of the sources in this merger remnant galaxy are like normal 1.4 \Msun\, accreting neutron star low mass X-ray binaries (LMXBs) and are shown by green circle. Four of the remaining are of high mass X-ray binary type (HMXB), four of heavily absorbed type, one of super-soft type and one of heavily absorbed very hard source, perhaps the  AGN. Though NGC 1316 is defined as star forming galaxy, we found only one super-soft source in this galaxy. In star forming galaxies, one expect relatively larger population of the super soft sources.
 
\section{Discussion}
\begin{figure}
\centering
\includegraphics[width=8cm,height=8cm]{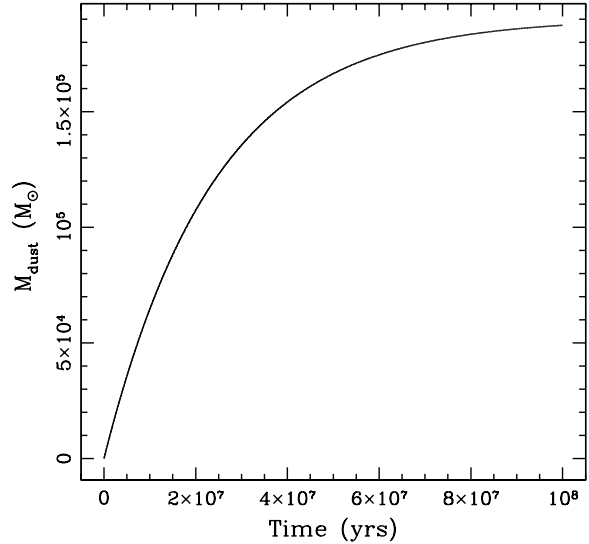} 
\caption{\label{accu} Total build-up of the dust within NGC 1316 considering two competitive processes i.e. injection versus destruction of the grains.}
\end{figure}

The issue of origin of dust and gas in early-type galaxies is highly controversial. Internal origin of the dust in this class of galaxies assumes contribution mainly from the asymptotic giant branch (AGB) stars. Supernovae (SNe) have also been recognized as a potential candidate for the injection of dust and gas in the ISM \citep{2007ApJ...662..927D}. The dust that is injected in to the ISM is simultaneously processed by the forward and reverse shocks in the hot gas swept up by SNe and hence may undergo rapid erosion by sputtering \citep{2010PhRvD..81h3007N}. With the observed SNe rate in NGC 1316 and considering the two competitive processes of formation and simultaneous destruction of the grains, one can estimate total content of dust that a galaxy may accumulate over its life time by solving the empirical relation (see \citealt{1999AJ....118..785D} for details), 
 \[\frac{\partial M_d(t)} {\partial t} = \frac{\partial M_{d,s}}{\partial t} - M_d(t)\,  \tau_d^{-1} \] 
where $\frac{\partial M_d(t)} {\partial t}$ is the net dust accumulation rate; $\frac{\partial M_{d,s}}{\partial t}$ is the rate at which dust is being injected by the SNe and stellar winds; M$_d(t)$ is the amount of dust available at a given time t; and $\tau_d^{-1}$ is the dust grain destruction rate. The life of grains of radius $a$ against sputtering due to electrons, hot protons and $\alpha$-particles can be estimated employing the relation given by \cite{1979ApJ...231..438D} and in the present case it is found to be equal to 4.1$\times\, 10^{-8}\, yr^{-1}$. By assuming gas-to-dust ratio about $\sim$ 100, the dust injection rate in NGC 1316 would be about 0.09 \Msun\, yr$^{-1}$. Therefore, the total build up (Figure~\ref{accu}) of dust content in NGC 1316 over its life $\sim$3 Gyr may be equal to 2.0 $\times\, 10^{5}$ \Msun. 

Comparison of the theoretically estimated mass of the dust with its true content derived using IR flux densities imply that the observed dust is far larger than expected to be accumulated by the galaxy over its life time. This discrepancy between the two estimates may enhance further if we include the amount of dust estimated using the observations at sub-mm wavelength \citep{1995A&A...295..317C}. Our estimate based on the integrated \textit{MIPS} flux densities leads to a value of 3.21$\times$10$^7$ \Msun\, and is roughly two orders of magnitude larger than the amount of dust acquired internally by the galaxy. Thus, the internal supply of dust by the mass loss from evolved stars and SNe is inefficient to account for the observed amount of dust in this galaxy and hence favours its external origin through a merger like event. \cite{2010ApJ...721.1702L} demonstrate its formation through the merging of late type galaxies with a supply of 2$-$4 $\times$10$^9$ \Msun\, gas; enough to account for the observed values of dust and gas in this galaxy. There are several other evidences that favours the formation of Fornax A through a merger like event. The most important evidence is the presence of shells around the main optical part of the galaxy. Further evidence comes from the surprising similarity in the morphological appearance of the dust, warm gas and hot gas. Internally produced gas and dust must follow the stellar light distribution rather than preferred planes. Moreover, given the erosion of dust grains by sputtering, one expects anti-correlation between the internally produced dust and the ionized gas. However, there are growing evidences of the positive correlation between the two components, strongly recommending the formation of the system through a merger like process. 

\section{conclusions}
We present multi-wavelength imagery of NGC 1316 with an objective to study the content of dust and other phase of ISM and its origin in the system. Results derived from this study are:
\begin{itemize}
\item Color-index maps as well as extinction maps derived for NGC 1316 delineates intricate morphology of the dust. This galaxy hosts a prominent dust lane oriented along optical minor axis in the inner region, which then takes an arc like shape at about 6 kpc. Apart from these main features, several other knots/clumpy features are also evident in this galaxy. Dust emission maps derived using \emph{Spitzer} observations at 8 $\mu$m exhibit even more complicated structure of the dust distribution.
\item Extinction curve derived over the range of optical ($B, V, R\, \& I$) to near-IR ($J, H\, \& Ks$) bands for the NGC 1316 follows closely the standard Galactic extinction law, implying that properties of dust grains in the merger remnant galaxy are identical to the canonical grains in the Milky Way.
\item Dust content of NGC 1316 estimated using optical extinction is about 2.13 $\times\, 10^5$ \Msun\, and is an order of magnitude shorter than that derived using the IRAS flux densities and two orders of magnitude shorter than that using integrated flux densities from \textit{MIPS} at 24$\mu$m, 70$\mu$m and 160$\mu$m.
\item Morphology of the hot gas derived from the analysis of high resolution \textit{Chandra} observations and ionized gas morphology mapped through H$_{\alpha}$  emission exhibits surprising similarities together with the  morphology of the dust, pointing towards the common origin of all the phases of ISM.
\item Combined spectrum of the point sources removed X-ray photons within optical D$_{25}$ region is well constrained by a double temperature together with the power law component, the power-law component confirming the contribution from the unresolved population of sources.
\item X-ray analysis enabled us to detect a total of 80 discrete sources within D$_{25}$, majority of which are the normal LMXBs. X-ray color-color plot of the resolved sources exhibited the structural differences of the sources.
\item Spatially resolved spectral analysis of the X-ray photons exhibited temperature structure showing a positive gradient in it as a function of radial distance.
\end{itemize}

\section*{Acknowledgments}
The authors are grateful to the anonymous referee for their careful reading and encouraging comments on the manuscript, that enabled us to  improve the quality of the paper. This work is supported by UGC, New-Delhi under the major research project F.No. (36-240/2008-SR). We acknowledge the use of High Performance Computing Facility developed under DST-FIST scheme sanction No. SR/FST/PSI-145. Usage of facilities at IUCAA, Pune are gratefully acknowledged. This work has made use of data from the NASA/IPAC Extragalactic Database (NED), which is operated by the Jet Propulsion Laboratory, California Institute of Technology, under contract with the National Aeronautics and Space Administration. This publication has made use of data products from the \textit{Chandra}, CTIO, GALEX, 2MASS, IRAS and MIPS archives. 


\def\aj{AJ}%
\def\actaa{Acta Astron.}%
\def\araa{ARA\&A}%
\def\apj{ApJ}%
\def\apjl{ApJ}%
\def\apjs{ApJS}%
\def\ao{Appl.~Opt.}%
\def\apss{Ap\&SS}%
\def\aap{A\&A}%
\def\aapr{A\&A~Rev.}%
\def\aaps{A\&AS}%
\def\azh{AZh}%
\def\baas{BAAS}%
\def\bac{Bull. astr. Inst. Czechosl.}%
\def\caa{Chinese Astron. Astrophys.}%
\def\cjaa{Chinese J. Astron. Astrophys.}%
\def\icarus{Icarus}%
\def\jcap{J. Cosmology Astropart. Phys.}%
\def\jrasc{JRASC}%
\def\mnras{MNRAS}%
\def\memras{MmRAS}%
\def\na{New A}%
\def\nar{New A Rev.}%
\def\pasa{PASA}%
\def\pra{Phys.~Rev.~A}%
\def\prb{Phys.~Rev.~B}%
\def\prc{Phys.~Rev.~C}%
\def\prd{Phys.~Rev.~D}%
\def\pre{Phys.~Rev.~E}%
\def\prl{Phys.~Rev.~Lett.}%
\def\pasp{PASP}%
\def\pasj{PASJ}%
\def\qjras{QJRAS}%
\def\rmxaa{Rev. Mexicana Astron. Astrofis.}%
\def\skytel{S\&T}%
\def\solphys{Sol.~Phys.}%
\def\sovast{Soviet~Ast.}%
\def\ssr{Space~Sci.~Rev.}%
\def\zap{ZAp}%
\def\nat{Nature}%
\def\iaucirc{IAU~Circ.}%
\def\aplett{Astrophys.~Lett.}%
\def\apspr{Astrophys.~Space~Phys.~Res.}%
\def\bain{Bull.~Astron.~Inst.~Netherlands}%
\def\fcp{Fund.~Cosmic~Phys.}%
\def\gca{Geochim.~Cosmochim.~Acta}%
\def\grl{Geophys.~Res.~Lett.}%
\def\jcp{J.~Chem.~Phys.}%
\def\jgr{J.~Geophys.~Res.}%
\def\jqsrt{J.~Quant.~Spec.~Radiat.~Transf.}%
\def\memsai{Mem.~Soc.~Astron.~Italiana}%
\def\nphysa{Nucl.~Phys.~A}%
\def\physrep{Phys.~Rep.}%
\def\physscr{Phys.~Scr}%
\def\planss{Planet.~Space~Sci.}%
\def\procspie{Proc.~SPIE}%
\let\astap=\aap
\let\apjlett=\apjl
\let\apjsupp=\apjs
\let\applopt=\ao
\bibliographystyle{raa}
\bibliography{mybib}
\end{document}